\documentclass[10pt]{iopart}
\begin{document}

\title[Radiation reaction]{Perturbation method in the assessment of radiation reaction in the capture of stars by black holes}
\author{Alessandro D.A.M. Spallicci\dag \footnote[3]{
Dep. d'Astrophysique Relativiste: Th\`eories, Exp\'eriences, Mesures, Instrumentations, Signaux. 
Boulevard de l'Observatoire, BP 4229, 06304 Nice France. Email spallicci@obs-nice.fr}, 
Sofiane Aoudia\dag}

\address{\dag\ Observatoire de la C\^ote d'Azur, Nice}

\begin{abstract}
This work deals with the motion of a radially falling star in Schwarzschild geometry and correctly identifies radiation reaction terms by the perturbative method. The results are: i) identification of all terms up to first order in perturbations, second in trajectory deviation, and mixed terms including lowest order radiation reaction terms; ii) renormalisation of all divergent terms by the $\zeta$ Riemann and Hurwitz functions. The work implements a method previously identified by one of the authors and corrects some current misconceptions and results.  
\end{abstract}


\section{Introduction}

The burst of gravitational waves emitted during the capture of a compact star by a supermassive black hole is of main interest for 
space interferometry \cite{lisa}, for testing general relativity in strong field and for investigating the physics of black holes. 
The detection of such sources requires the design of templates, which in turns require understanding of the complex orbital evolution 
during capture. These considerations are timely and strategical, especially if we refer to the supermassive black hole in the centre of our galaxy \cite{ghezetal00}, as a source to be potentially detected in the next years.\\     
In the last years, perturbation methods have played an increasingly important role, due to their applicability to the 
last phases of lifetime of binaries constituted by comparable masses before merge \cite{pripul94}. The implication for ground 
interferometry \cite{vili} has not passed unnoticed \cite{bakcamlou02} and considerable efforts and resources have been and are 
being employed. \\   
The understanding of orbital evolution requires understanding of radiation reaction effects. Indeed, radiation reaction, 
still partially outstanding problem in general relativity, like the two-body problem, is of most concern for gravitational waves detectors. Its  influence is manifest in data analysis, where a phase mismatch of the templates with the signals may cause loss of detection. \\
In this work, we analyse a radially falling star, m, captured by a massive black hole, M, by perturbative methods.
The motion is studied in strong gravity, the perturbation being based on the $m/M$ ratio. Radial fall is an idealisation of the capture scenario, but applicable to final plunging. Furthermore, most of the radiation, and thus reaction, occurs close to the horizon where 
inspiral has ceased. Nevertheless, we do not make specific reference to plunging as our analysis allows any distance 
for the point of departure.\\
We refrain from using energy balance and adiabatic hypothesis. The former is the imposition at start, rather than a  rightful outcome, of the equality 
of the energy radiated with the energy loosed by the system. The latter can't be evoked since the particle immediately has to 
react to the radiation emitted, contrarily to inspiral motion where radiation reaction time scale is larger than the orbital period\footnote{Inspiral motion, especially around spinning black holes, at high eccentricity and on inclined orbits is characterised by three radiation time scales \cite{hug01}.}. Furthermore, an analysis out of adiabacity implies that a straightforward linearisation of the phenomenon under study is not justified, and instead a careful screening of relevant terms must be adopted. Hence, neglection of higher order terms is to be avoided in absence of compelling reasons. \\    
We study the motion of a radially falling star with a perturbative approach using Moncrief \cite{mon74} gauge invariant formulation of the Regge-Wheeler \cite{rw57} and Zerilli \cite{zeri} equations. Incidentally, also second order perturbations formalism may be made gauge invariant \cite{garpri00}. Under these assumptions, we shall not be concerned with gauge issues hereafter.\\
Finally, second order perturbation analysis \cite{glniprpu96} demands a consistent description at first order of the energy momentum tensor. The linearisation of general relativity implies that the particle motion in unperturbed Schwarzschild generates radiation; such radiation, including quadratic radiative terms of first order,  plus the corrected motion, including radiation reaction, shall generate radiation at second order.  Computing gravitational radiation to second order 
requires the knowledge of the 
trajectory of the falling mass on the first order metric (Schwarzschild plus perturbations).    \\
In the second section, it shall accordingly be developed a physical hierarchical scheme for each term constituting the geodesic, classifying terms that produce accelerations deviations from the unperturbed geodesic in Schwarzschild geometry
in three types: the first depending upon the unperturbed metric evaluated on the {\it perturbed} trajectory; the second 
upon the {\it perturbed} metric evaluated on the unperturbed trajectory; the third upon the {\it perturbed} metric evaluated on the {\it perturbed} trajectory. In the third section, the issue of renormalisation shall be dealt for the infinite sum, on all multipoles, of finite terms that lead to divergencies. \\
We adopt the geodesic concept in our approach, meaning that any motion is geodesic if the underlying metric is properly defined.\\ 
Finally we are revising the work on the concept of self-force \cite{sf} and correspondences between the two methods 
are subject of an undergoing investigation \cite{as}.

\section{The metric, the perturbation scheme and the geodesic equation}

Perturbation method for analysis of radiation reaction has been previously proposed \cite{spa99a} - \cite{spa00}.
The metric is the sum of the Schwarzschild metric and the perturbations:

\[
{\fl
\eta_{\mu\nu} \!=\! 
\pmatrix{ f & 0\cr
	0 & - \displaystyle\frac{1}{f}\cr}
~
h_{\mu\nu}\!=\! 
\pmatrix{ - f H_0 & - H_1\cr
	- H_1 & - \displaystyle\frac{1}{f}H_2\cr}
~
\eta^{\mu\nu}\!=\! 
\pmatrix{ \displaystyle\frac{1}{f} & 0\cr
	0 & - f \cr}
~
h^{\mu\nu}\!=\! 
\pmatrix{ - \displaystyle\frac{1}{f}H_0 & H_1\cr
	H_1 & - fH_2\cr}
}
\]

\[
g_{tt} = f(1 - H_0)~~~~g_{tr} = g_{rt} = - H_1~~~g_{rr} = - \displaystyle\frac{1}{f}(1 + H_2)
\]
\[
g^{tt} = \displaystyle\frac{1}{f}(1 + H_0)~~~~g^{tr} = g^{rt} = - H_1~~~~g^{rr} = - f(1 - H_2)
\]
\[
g_{\mu\nu} = \eta_{\mu\nu} + h_{\mu\nu}~~~~~~~~~g^{\mu\nu} = \eta^{\mu\nu} - h^{\mu\nu}
~~~~~~~~~f = \frac{r - 2 M}{r}
\]

The order of the perturbation can be made explicit, i.e. $h^{(1)},h^{(2)}$. The position of the particle $r_e = r_p + \Delta r_p$ is given by the unperturbed trajectory in the unperturbed field $r_p$ and 
by several contributions, among which radiation reaction, given by the unperturbed and perturbed field, that generate a trajectory deviation $\Delta r_p$. 
The field is developed in Taylor series around the {\it real} position of the particle: $g_{\mu\nu}(r_e) = g_{\mu\nu}(r_p) + 
\Delta r_p \left (\partial g_{\mu\nu}/\partial r \right )_{r_p}$. The geodesic is 
only dependent upon radial and time coordinates:

\begin{equation}
{\fl 
\frac{d^2 r}{dt^2} = \Gamma_{rr}^t \left(\frac{dr}{dt}\right)^3 + \left(2\Gamma_{tr}^t - \Gamma_{rr}^r\right)\left(\frac{dr}{dt}\right)^2
+ \left(\Gamma_{tt}^t - 2 \Gamma_{tr}^r\right)\left(\frac{dr}{dt}\right) 
- \Gamma_{tt}^r 
\label{eq:geolou}
 }
\end{equation}

In absence of the weak field hypothesis, $h$ is not limited in amplitude, but the following justifies that solely the terms in {\it Tab. 1} are to be retained. We suppose: 

\begin{equation}
\frac{[h^{(1)}]^2}{\eta}  \simeq \frac{h^{(2)}}{\eta} \ll
\frac{h^{(1)}}{\eta} 
<
\frac{\Delta {\ddot r}_p}{{\ddot r}_p}
\label{eq:hyp}
\end{equation}

The latter inequivalence is due to the twofold nature of the acceleration trajectory deviation: $\Delta {\ddot r}_p$ is the sum of two types of contributions. One is given by the Schwarzschild metric\footnote{Indeed, the first five terms $\alpha_{1-5}$ provide an acceleration 
${\ddot r}_p$ depending solely upon $\eta$, the unperturbed metric}, the other by the perturbations $h$ (coupled with $\eta$):

\begin{equation}
\Delta {\ddot r}_p = \Delta {\ddot r}_p(\eta) + \Delta {\ddot r}_p(h)
\end{equation}

The acceleration is dependent upon $h^{(1)}$ derivatives which are not necessarily small, especially in 
the last phase of the trajectory. In conclusion, the terms proportional to $h^{(1)}, {\Delta r_p}, {\Delta {\dot r}_p}$, $h^{(1)}$ derivatives
 and ${\Delta r_p^2}, {\Delta {\dot r}_p^2}, {\Delta r_p}{\Delta{\dot r}_p}$ are retained, while those to $[h^{(1)}]^2$, and 
$h^{(2)}$ are neglected, as the second order terms in trajectory deviation when multiplied by first order perturbations.

We write the geodesic equation in the following form:

\begin{equation}
\fl
\Delta {\ddot r}_p = \alpha_1 \Delta r_p + \alpha_2 \Delta {\dot r}_p + \alpha_3 \Delta r_p^2  
+ \alpha_4 \Delta {\dot r}_p^2 + \alpha_5 \Delta r_p \Delta {\dot r}_p + \alpha_6 +
\alpha_7 \Delta r_p + \alpha_8 \Delta {\dot r}_p
\label{eq:geoalpha}
\end{equation} 

The physical significance of the terms is essential\footnote{The $\alpha_1$ term correspond to the A term of 
\cite{lo00},\cite{lo01a}, apart of an error in the quoted publications (see appendix). 
The $\alpha_2$ term correspond to the B term, $\alpha_6$ correspond to C of \cite{lo00},\cite{lo01a}.}.
The terms $\alpha_{1,2,3,4,5}$ arise from the pure Schwarzschild metric and they may be alternatively interpreted as representing the coefficients of the geodesic deviation of two particles separated on the radial axis by a $\Delta r_p$ distance. In the 
scenario of a single falling particle, they are the coefficients representing the {\it unperturbed} Schwarzschild metric influence calculated on the {\it perturbed } particle trajectory, i.e. the {\it real} position ($\alpha_{1,3}$), velocity ($\alpha_{2,4}$) or both ($\alpha_{5}$).         
The $\alpha_6$ term is the lowest order containing the perturbations. It represents the perturbation influence calculated at the position and velocity of the particle in the {\it unperturbed} trajectory. Thus it does {\it not} represent radiation reaction, although it contributes to.
The $\alpha_7$ term represents the perturbation influence calculated on the {\it real} position of the particle in the {\it perturbed} trajectory, $h_{\mu\nu} \Delta r_p$. Finally, the $\alpha_8$ term represents the perturbation influence calculated on the {\it real} velocity of the particle in the {\it perturbed} trajectory $h_{\mu\nu} \Delta \dot{r}_p$.  \\
It is thus a development at first order in perturbations and second order in trajectory deviation or mixed terms. 
{\it Tab. 1} lists all $\alpha$ terms.
We emphasize that:
\begin{itemize}
\item{The terms $\alpha_{3,4,5,7,8}$ were previously\cite{lo00},\cite{lo01a} neglected and the remaining terms 
$\alpha_{1,2,6}$ do not really represent radiation reaction; ii) the terms $\alpha_{3,4,5}$ are of second order     
in trajectory deviation. They also do not represent radiation reaction but are kept for mathematical consistency with 
the hypothesis (\ref{eq:hyp})\footnote{It shall be the upcoming numerical simulation to verify the relative weight of these 
terms \cite{vin04}.}.
The terms $\alpha_{7,8}$ represent the lowest order radiation reaction terms.}
\item{Solely terms in second order of perturbations are excluded. Such choice is also compliant to the nature of the energy-momentum tensor 
of the second order equations. The tensor $T_{\mu\nu}$, that generates second order perturbations, must entail a geodesic 
stemmed from eq.(\ref{eq:geoalpha}). The linearisation of general relativity induces a stepped-up approach, in where radiation and motion 
assume sequentially the lead. Motion generates radiation that generates reaction, which in turns generates radiation of second order (to be dealt with squared first order radiation terms). But radiation of second order requires generation by radiation reaction at first(lowest) order, i.e. the terms $\alpha_{7,8}$. }
\item{Radiation reaction effects, supposedly,  are larger near the horizon, where most of the radiation is emitted. The leading terms of  
$\alpha_{6}$ near the horizon are of the order
$h^{rr}\eta_{rr,r}\dot{r}_p^2 + \eta^{tt}h_{tt,t}\dot{r}_p + h^{rr}\eta_{tt,r}$, whereas in $\alpha_{7,8}$, are of the order 
$h^{rr}\eta_{rr,rr}\dot{r}_p^2 + \eta^{tt}_{,r}h_{tt,t}\dot{r}_p$ and
$h^{rr}\eta_{rr,r}\dot{r}_p + \eta^{tt}h_{tt,t}$ to be multiplied by 
$\Delta r_p$ and $\Delta \dot{r}_p$, respectively. The Schwarzschild metric tensor components and derivatives, above quoted, tend to infinity and its powers, while $\dot{r}_p$ to zero, near the horizon. Estimate of the leading terms in $\alpha_{7,8}$ justifies their inclusion.}
\item{Second order perturbations equations, are yet unsolved in presence of a source term and the vacuum solutions are adequate for dealing the ``close limit'' but not the particle motion. The unavailability of the second order perturbation solutions suggest to 
consider the problem pragmatically using all remaining known quantities, as shown in eq.(\ref{eq:geoalpha}).}

\end{itemize}

\section{Black hole polar perturbations equation }

Zerilli \cite{zeri}
found the equation for polar perturbations and studied the 
emitted radiation adding a source term, 
a freely falling test mass $m$ into the black $M$. The equation is written in terms of the 
wavefunction $\psi_l$ for each l-pole component, the tortoise coordinate $r^*$, the polar potential $V_{l}(r)$, 
the $2^{l}$-pole source component $S_{l}(r,t)$:

\begin{equation}
\fl
\frac {d^{2} \psi_l(r,t)}{dr^{*2}} - \frac {d^{2} \psi_l(r,t)}{dt^2} -
V_{l}(r)\psi_l (r,t) = S_{l}(r,t)  
~~~~~~~~~
r^{*}= r + 2M \ln \left ({\displaystyle{\frac{r}{2M}}} - 1 \right )
\end{equation}

\begin{equation}
\fl
V_{l}(r) = \left ( 1- \frac{2M}{r}\right ) \frac {2\lambda^{2}(\lambda +
1)r^{3} + 6\lambda^{2}Mr^{2} + 18\lambda M^{2}r + 18M^{3}} {r^{3}(\lambda r
+ 3M)^2}
~~~\lambda = {\frac{1}{2}} (l - 1) (l + 2)
\end{equation}

\[
S_{l} = \frac{\left( 1-{\displaystyle\frac{2M}{r}}\right) 4M\sqrt{(2l+1)\pi }}{(\lambda
+1)(\lambda r+3M)} \times
\]
\begin{equation}
\fl
\left \{r\left( 1-{\displaystyle\frac{2M}{r}}\right) ^{2}\delta
^{\prime }[r-r_{p}(t)]
- \left ( \lambda +1 -\frac{M}{r}-\frac{6Mr}{\lambda r+3M}\right ) \delta
[r-r_{p}(t)] \right \}
\end{equation}

where $r_{p}(t)$, geodesic in {\it unperturbed Schwarzschild metric}, is the inverse of:

\begin{equation}
\fl
t=-4M\left( {\frac{r}{2M}}\right) ^{1/2}-\frac{4M}{3}\left( \frac{r}{2M}
\right) ^{3/2}-2M\ln \left [\left( \sqrt{\displaystyle{\frac{r}{2M}}}-1 \right) \left(
\sqrt{\displaystyle{\frac{r}{2M}}}
+1\right )^{-1}\right]   \label{eq:tofr}
\end{equation}
The perturbations around the particle are (Regge-Wheeler gauge $H_0^l = H_2^l$):

\[ 
\fl
H_0^l = - \frac{9 M^3 + 9 \lambda M^2 r + 3 \lambda ^2 M r^2 + \lambda ^2(\lambda + 1)r^3}{r^2(\lambda r + 3M)^2} \psi
+ 
\]
\begin{equation}
\frac{3 M^2 - \lambda M r + \lambda r^2 }{r(\lambda r + 3M)} \psi_{,r}
+ (r - 2M)  \psi_{,rr}
\label{eq:h2topsi}
\end{equation}

\begin{equation}
H_1^l = r \psi_{,rt} - \frac{3 M^2 + 3 \lambda M r -\lambda r^2}{(r - 2 M)(\lambda r + 3M)^2} \psi_{,t}
\label{eq:h1topsi}
\end{equation}

The {\it unperturbed} velocity is given by ($r_0$ is the test mass position at start):

\begin{equation}
{\dot r}_p = - \left( 1 - \frac{2M}{r_p}\right)
\left( \frac{2M}{r_p} - \frac{2M}{r_0}\right)^{1/2}
\left(1 - \frac{2M}{r_0}\right)^{- 1/2}
\end{equation}

\section{Renormalisation}

In this section we reconstruct the renormalisation for $\alpha_6$ and apply, for the first time, the 
$\zeta$ function to  $\alpha_{7,8}$. Indeed, the infinite sum over the finite multipole components contributions leads to the problem of dealing 
infinities in the results. For ever larger $l$ the metric perturbations tend to an asymptotic behaviour. In other words, the curves representing each metric perturbation component for each l, accumulate over the $l\rightarrow \infty$ curve. Thus the subtraction from each mode of the $l\rightarrow \infty$ leads to a convergent series. 
We extend the application of 
the Riemann $\zeta$ function for renormalisation \cite{lo00} - \cite{lo01a} to all pertinent terms of the 
geodesic of {\it Tab. 1}. Instead, mode-sum 
renormalisation is planned in the near future.
For $L = l + 0.5$, the wavefunction and its derivatives assume the following forms at large $L$ or $l$ \cite{lo01a}, 
\cite{lo01b}\footnote{The derivation of such expressions, quoted from a paper in preparation by Barack and Lousto, as  
referred by \cite{lo01a} and \cite{balo02}, is yet unpublished.} when 
averaged around the particle at $r_p$:

\[
\fl
\bar{\psi} \simeq 4 \sqrt{2\pi}mL^{-2.5} 
~~~~~~~
\bar{\psi}_{,r} \simeq - \frac{6 \sqrt{2\pi}m(r_0 - 2M)}{r_0(r_p - 2M)} L^{-2.5} 
~~~~~~~
\bar{\psi}_{,rr} \simeq 
 \frac{4 \sqrt{2\pi}m(r_0 - 2M)}{r_0(r_p - 2M)^2} L^{-0.5} 
\] 

\[
\bar{\psi}_{,rrr}  \simeq \frac{4 \sqrt{2\pi}m (r_0 - 2M)}{r_0(r_p - 2M)^3}
\left[\frac{5(r_0 - 2M)}{2 r_0} + \frac{9M}{r_p} - 6 \right] L^{-0.5} 
\] 

\[
\fl
\bar{\psi}_{,t} \simeq  
\frac{6 \sqrt{2\pi}m \sqrt{r_0 - 2 M}\dot{r}_p}{\sqrt{r_0} r_p} L^{-2.5} 
~~~~~~~~~~~~~~
\bar\psi_{,tr} \simeq -\frac{4 \sqrt{2\pi}m \sqrt{r_0 - 2 M} \dot{r_p} }{\sqrt{r_0}r_p(r_p - 2M) }
L^{-0.5}   
\]

\[
\bar{\psi}_{,trr}  =  - \frac{4 \sqrt{2\pi}m \sqrt{r_0 - 2 M} \dot{r_p} }{\sqrt{r_0}r_p(r_p - 2M)^2 }
\left[\frac{5 (r_0 - 2 M)}{2 r_0 } + \frac{9 M}{r_p}- 4 \right ]L^{-0.5} 
\]

Using the above equations, eqs.(\ref{eq:h2topsi},\ref{eq:h1topsi}) and recasting $H_{1,t}$ as function of $
\bar{\psi},~ 
\bar{\psi}_{,r},~ 
\bar{\psi}_{,rr},~  
\bar{\psi}_{,rrr}$, the $\alpha_6$ term for large $L$ or $l$ around the particle is:

\begin{equation}
\alpha_6 = \sum_{l=0}^\infty \alpha_6^l
~~~~~~~~~~~~~
\alpha_6^l = \alpha_6^a L^0 + \alpha_6^b L^{-2} + \alpha_6^c L^{-4} + O(L^{-6})
\label{eq:cdanorm}
\end{equation}

The term $\alpha_6^a L^0$ needs\footnote{The term $\alpha_6^a L^0$ differs from b of eq.(13) in \cite{lo01a} which is again different from the value given in \cite{lo01b}.} renormalisation \cite{ao03}. 
The Riemann $\zeta$ function \cite{ri59} and its generalisation, the Hurwitz $\zeta$ function \cite{hu82}, are defined by:

\begin{equation}
\zeta (s) = \sum_{l=1}^\infty (l)^{-s}
~~~~~~~~~~~~~
\zeta (s,a) = \sum_{l=0}^\infty (l + a)^{-s}
\end{equation}

where in our case $a = 0.5$. Thus:

\begin{equation}
\zeta (s, 0.5) = \sum_{l=0}^\infty \left (l + 0.5 \right )^{-s} = 2^s \left[\sum_{l=0}^\infty (2l + 1)^{-s}\right]
\label{eq:191}
\end{equation}    

Due to the imparity of the term in braces, eq.(\ref{eq:191})is rewritten as:

\begin{equation}
\fl
\zeta (s, 0.5) = 2^s \left \{\zeta(s) - \left[\sum_{l=0}^\infty (2l)^{-s}\right]\right\} = 2^s \left ( 1 - 2^{-s} \right)
\zeta(s) =     \left (2^{s} - 1 \right)\zeta(s) 
\label{eq:192}
\end{equation}    
 
Some special values of the Hurwitz functions are:

\begin{equation}
\fl
\zeta (-2, 0.5) = 0~~~~~~~~~~~~\zeta (0, 0.5) = 0~~~~~~~~~~~~\zeta (2, 0.5) = \frac{1}{2}\pi^2~~~~~~~~~~~~\zeta (4, 0.5) = 
\frac{1}{6}\pi^4
\end{equation}

The latter values when applied to eq.(\ref{eq:cdanorm}), give:

\[
\fl
\alpha_6 = \alpha_6^a\sum_{l=0}^\infty (l + 0.5)^0 + \alpha_6^b\sum_{l=0}^\infty (l + 0.5)^{-2} 
+ \alpha_6^c\sum_{l=0}^\infty (l + 0.5)^{-4} + [0(l + 0.5)^{-6}]=
\] 
\begin{equation}
\fl
\alpha_6^a\zeta(0,0.5) + \alpha_6^b\zeta(2,0.5) + \alpha_6^c\zeta(4,0.5)+ [0(l + 0.5)^{-6}]
= \frac{1}{2}\pi^2\alpha_6^b + \frac{1}{6}\pi^4\alpha_6^c + [0(l + 0.5)^{-6}]
\end{equation}

For the renormalisation of $\alpha_{7,8}$ terms, the expressions: 
$
\bar{\psi}_{,tt} 
\bar{\psi}_{,ttt}
\bar{\psi}_{,ttr}
\bar{\psi}_{,trrr} 
\bar{\psi}_{,rrrr}      
$
are deducted \cite{ao03} operating on the averaged wavefunctions and derivatives, and the homogeneous wave 
equation. The latter is recast as \cite{spa99a} - \cite{spa99b}:

\begin{equation}
\frac{1}{\rho^2}\frac {d^{2} \psi_l(r,t)}{dr^{2}} - \frac {d^{2} \psi_l(r,t)}{dt^2} + \frac{\rho - 1}{r\rho^2}\frac {d \psi_l(r,t)}{dr} 
- V_{l}(r)\psi_l (r,t) = 0
\label{eq:rwzh}
\end{equation}

where $\rho = dr^*/dr$.  
Deriving sequentially eq.(\ref{eq:rwzh}), we get the $\psi$ needed derivatives. 
The latter are evaluated for $L\rightarrow\infty$ and when inserted in the $\alpha_{7,8}$ terms, result into:

\begin{equation}
\fl
\alpha_7 = \sum_{l=0}^\infty \alpha_7^l
~~~~~~~~~~~~~~
\alpha_7^l =  \alpha_7^a L^2 + \alpha_7^b L^0 + \alpha_7^c L^{-2} + \alpha_7^d L^{-4} + O(L^{-6})
\label{eq:7danorm}
\end{equation}

\begin{equation}
\fl
\alpha_8 = \sum_{l=0}^\infty \alpha_8^l
~~~~~~~~~~~~~~~~~~~~~~
\alpha_8^l = \alpha_8^a L^0 + \alpha_8^b L^{-2} + \alpha_8^c L^{-4} + O(L^{-6})
\label{eq:8danorm}
\end{equation}

Renormalisation of eqs. (\ref{eq:7danorm},\ref{eq:8danorm}) leads to:

\begin{equation}
\fl
\alpha_7 = \frac{1}{2}\pi^2\alpha_7^c + \frac{1}{6}\pi^4\alpha_7^d + [0(l + 0.5)^{-6}]
~~~~~~
\alpha_8 = \frac{1}{2}\pi^2\alpha_8^b + \frac{1}{6}\pi^4\alpha_8^c + [0(l + 0.5)^{-6}]
\end{equation}

\section{Conclusions}

We have obtained the following results: i) calculation and determination of all terms up to first order in perturbations, second in trajectory deviation, mixed term of second order including lowest order radiation reaction terms, all contributing 
to the trajectory of a radially falling test  mass in Schwarzschild geometry; ii)
renormalisation of all divergent terms, including new ones, stemmed from the infinite sum of finite angular momentum dependent components by the $\zeta$ Riemann and Hurwitz functions; iii) correction and improvements of previously published results.

\section{Acknowledgments}

AS wishes to thank G. Sch\"{a}fer (Jena) for an illuminating discussion on radiation reaction held at the $2^{nd}$ Amaldi, 
B. Chauvineau (Grasse) for support in the analysis 
of the geodesic in \cite{lo00},\cite{lo01a}.  This research work was supported by ESA 
with a G. Colombo Senior Research Fellowship to A. Spallicci.

\section*{Appendix on the geodesic equation}

Three perturbations schemes concur: 

\begin{equation}
\fl
r_e = r_p + \Delta r_p
~~~~
g_{\mu\nu}(r_e) = g_{\mu\nu}(r_p) + 
\Delta r_p \left(\frac{\partial g_{\mu\nu}}{\partial r}\right )_{r_p}
~~~~
g_{\mu\nu} = \eta_{\mu\nu} + h_{\mu\nu}
\end{equation} 

Hence. the development of eq.(\ref{eq:geolou}) leads to a lengthy computation

\[
\Gamma_{rr}^t \left(\frac{dr}{dt}\right)^3  \simeq 
\frac{1}{2}\left [g^{tt}(2 g_{tr,r} - g_{rr,t}) + g^{tr}g_{rr,r}\right ]
(\dot{r}_p^3 + 3 \dot{r}_p^2 \Delta\dot{r}_p + 3 \dot{r}_p \Delta\dot{r}_p^2) \simeq
\]
\[
\fl
\frac{1}{2}
\left [
\left(\eta^{tt} + h^{tt} + \eta^{tt}_{,r} \Delta r_p + h^{tt}_{,r} \Delta r_p\right)
\left(2 h_{tr,r} + 2 h_{tr,rr} \Delta r_p - h_{rr,t} - h_{rr,tr} \Delta r_p \right)
+ \right.
\]
\begin{equation}
\fl
\left.
\left(h^{tr} + h^{tr}_{,r} \Delta r_p \right)
\left(\eta_{rr,r} + h_{rr,r} + \eta_{rr,rr}\Delta r_p + h_{rr,rr}\Delta r_p \right)
\right ] 
(\dot{r}_p^3 + 3 \dot{r}_p^2 \Delta\dot{r}_p + 3 \dot{r}_p \Delta\dot{r}_p^2)
\label{eq:gamrrt}
\end{equation}
 
\[
2\Gamma_{tr}^t \left(\frac{dr}{dt}\right)^2 =
\left [g^{tt}g_{tt,r} + g^{tr}g_{rr,t}\right ]
(\dot{r}_p^2 + 2 \dot{r}_p \Delta\dot{r}_p + \Delta\dot{r}_p^2) 
\simeq 
\]
\[
\fl
\left [
\left(\eta^{tt} + h^{tt} + \eta^{tt}_{,r} \Delta r_p + h^{tt}_{,r} \Delta r_p\right)
\left(\eta_{tt,r} + h_{tt,r} +  \eta_{tt,rr}\Delta r_p + h_{tt,rr} \Delta r_p \right)
+ \right.
\] 
\begin{equation}
\fl
\left.
\left(h^{tr} + h^{tr}_{,r} \Delta r_p \right)
\left( h_{rr,t} + h_{rr,tr}\Delta r_p \right)
\right ] 
(\dot{r}_p^2 + 2 \dot{r}_p \Delta\dot{r}_p + \Delta\dot{r}_p^2 ) 
\label{eq:gamtrt}
\end{equation}

\[
- \Gamma_{rr}^r \left(\frac{dr}{dt}\right)^2 =
- \frac{1}{2} \left [g^{rr}g_{rr,r} + g^{rt}\left(2g_{tr,r} - g_{rr,t} \right )\right ]
(\dot{r}_p^2 + 2 \dot{r}_p \Delta\dot{r}_p + \Delta\dot{r}_p^2) 
\]
\[
\fl
\simeq 
- \frac{1}{2}
\left [
\left(\eta^{rr} + h^{rr} + \eta^{rr}_{,r} \Delta r_p + h^{rr}_{,r} \Delta r_p\right)
\left(\eta_{rr,r} + h_{rr,r} +  \eta_{rr,rr}\Delta r_p + h_{rr,rr} \Delta r_p \right)
+ \right.
\] 
\begin{equation}
\fl
\left.
\left(h^{rt} + h^{rt}_{,r} \Delta r_p \right)
\left( 2 h_{tr,r} + 2 h_{tr,rr}\Delta r_p - h_{rr,t} - h_{rr,tr}\Delta r_p \right)
\right ] 
(\dot{r}_p^2 + 2 \dot{r}_p \Delta\dot{r}_p + \Delta\dot{r}_p^2 ) 
\label{eq:gamrrr}
\end{equation}

\[
\Gamma_{tt}^t \left(\frac{dr}{dt}\right) =
\frac{1}{2}\left[g^{tt}g_{tt,t}  + g^{tr}\left(2g_{rt,t} - g_{tt,r}\right )\right ] 
(\dot{r}_p + \Delta\dot{r}_p) 
\simeq 
\]
\begin{equation}
\fl
\frac{1}{2}
\left [
\left(\eta^{tt} + h^{tt} + \eta^{tt}_{,r} \Delta r_p + h^{tt}_{,r} \Delta r_p\right)
\left(h_{tt,t} + h_{tt,rr}\Delta r_p \right)
+ \right.
\end{equation}
\[
\fl
\left.
\left(h^{tr} + h^{tr}_{,r} \Delta r_p \right)
\left( 2 h_{rt,t} + 2 h_{rt,tr}\Delta r_p - \eta_{tt,r}  - h_{tt,r}- \eta_{tt,rr}\Delta r_p - h_{tt,rr}\Delta r_p\right)
\right ] 
(\dot{r}_p + \Delta\dot{r}_p) 
\label{eq:gamttt}
\]

\[
- 2\Gamma_{tr}^r \left(\frac{dr}{dt}\right) =
- \left[g^{rr}g_{rr,t} + g^{rt}g_{tt,r}\right]
(\dot{r}_p + \Delta\dot{r}_p) 
\simeq 
\]
\[
\fl
- \left [
\left(\eta^{rr} + h^{rr} + \eta^{rr}_{,r} \Delta r_p + h^{rr}_{,r} \Delta r_p\right)
\left(h_{rr,t} + h_{rr,tr}\Delta r_p \right)
+ \right.
\] 
\begin{equation}
\fl
\left.
\left(h^{rt} + h^{rt}_{,r} \Delta r_p \right)
\left(\eta_{tt,r} + h_{tt,r} + \eta_{tt,rr}\Delta r_p + h_{tt,rr}\Delta r_p\right)
\right ] 
(\dot{r}_p + \Delta\dot{r}_p) 
\label{eq:gamtrr}
\end{equation}

\[
- \Gamma_{tt}^r = - \frac{1}{2}\left[g^{rr}\left(2g_{rt,t} - g_{tt,r}\right ) + g^{rt}g_{tt,t} \right  ] 
\simeq
\]
\[
\fl
- \frac{1}{2}
\left [
\left(\eta^{rr} + h^{rr} + \eta^{rr}_{,r} \Delta r_p + h^{rr}_{,r} \Delta r_p\right)
\left(2h_{rt,t} + 2 h_{rt,tr}\Delta r_p - \eta_{tt,r} -  h_{tt,r} 
- \eta_{tt,rr}\Delta r_p -  h_{tt,rr}\Delta r_p
\right)
\right.
\] 
\begin{equation}
\fl
\left.
+ \left(h^{rt} + h^{rt}_{,r} \Delta r_p \right)
\left(h_{tt,t} + h_{tt,rr}\Delta r_p\right)
\right ] 
\label{eq:gamttr}
\end{equation}

The development of eqs.(\ref{eq:gamrrt} - \ref{eq:gamttr}) produces the coefficients vertically listed 
in {\it Tab. 1}. The lines, hierarchically developed in horizontal rows, show the $\alpha_{i}$ (i = 1,..8) terms of eq.(\ref{eq:geoalpha}).     
The e.g. $\alpha_{1}$ term is given by:  

\begin{equation}
\fl
 \alpha_{1} = \eta^{tt}_{,r}\eta_{tt,r}\dot{r}_p^2 + \eta^{tt}\eta_{tt,rr}\dot{r}_p^2 
- \frac{1}{2}\eta^{rr}_{,r}\eta_{rr,r}\dot{r}_p^2 - \frac{1}{2}\eta^{rr}\eta_{rr,rr}\dot{r}_p^2 
+ \frac{1}{2}\eta^{rr}_{,r}\eta_{tt,r} +  
\frac{1}{2}\eta^{rr}\eta_{tt,rr}
\end {equation}

and results into eq.(\ref{eq:chau}). It's obvious that a similar development 
takes place for all $\alpha$ terms. But for the numerical estimate of $\alpha_{6,7,8}$
terms, simulation of the radial fall in time domain is required \cite{vin04}. The simulation is to provide wavefunctions 
and thus metric perturbations.  Here below all $\alpha$ terms are shown (Regge-Wheeler gauge $H_0 = H_2$):

\begin{equation}
\alpha_0 = - \frac{M}{r} \left(\frac{r - 2 M}{r^2} - \frac{3M}{r - 2M}\dot{r}_p^2\right )
\end{equation}

\begin{equation}
\alpha_1 = - \frac{2M}{r^2}
\left[\frac{3M}{r^2} - \frac{1}{r}+ \frac{3(r - M)}{(r - 2M)^2}\dot{r}_p^2
\right ]
\label{eq:chau}
\end{equation} 

\begin{equation}
\alpha_2 = \frac{6M}{r(r - 2M)}\dot{r}_p^2
\end{equation} 

\begin{equation}
\alpha_3 =  \frac{4M^2}{r^3}
\left[\frac{1}{r} + \frac{r - 4M}{(r - 2M)^3}\dot{r}_p^2
\right ]
\end{equation} 

\begin{equation}
\alpha_4 = \frac{3M}{r(r - 2M)}
\end{equation} 

\begin{equation}
\alpha_5 =  -\frac{4M}{r (r - 2M)}
\left[\frac{M}{r(r - 2M)} + \frac{2}{r} + \frac{1}{r - 2M}\right ]\dot{r}_p
\end{equation} 

\[
\alpha_6 =  \frac{1}{r - 2M}
\left[\frac{r^2}{2(r - 2M)} {\dot H}_0 - \frac{M}{r - 2 M} H_1 - r H'_1 \right ]\dot{r}_p^3
- \frac{3}{2} H'_0 \dot{r}_p^2 - 3 \left( \frac{1}{2}{\dot H}_0 - \frac{M}{r^2} H_1 \right ) \dot{r}_p 
\]
\begin{equation}
+ \frac{r - 2M}{r}\left( \frac{2M}{r^2}H_0 - \frac{1}{2}\frac{r - 2M}{r} H'_0 - {\dot H}_1 \right ) 
\end{equation} 

\[
\alpha_7 =  - \frac{1}{r - 2M}
\left[\frac{2Mr}{(r - 2M)^2} {\dot H}_0 - \frac{1}{2}\frac{r^2}{r - 2M} {\dot H}'_0 - \frac{2M}{(r-2M)^2}H_1 
- \frac{M}{r-2M}H'_1 + r H''_1 \right ] \dot{r}_p^3  
\]
\begin{equation}
- \frac{3}{2} H''_0 \dot{r}_p^2
- \frac{3}{2}\left ({\dot H}'_0 + \frac{4M}{r^3}H_1 - \frac{2M}{r^2}H'_1 \right ) \dot{r}_p
\end{equation}
\[
- \frac{1}{r}
\left[\frac{4M (r - 3M)}{r^3} H_0 - \frac{4M(r - 2M)}{r^2}H'_0 - \frac{1}{2} \frac{(r - 2M)^2}{r}H''_0 
+ \frac{2M}{r}{\dot H}_1 + (r - 2M) {\dot H}'_1  \right ]
\] 

\[
\alpha_8 =  \frac{3}{r - 2M}
\left[\frac{1}{2} \frac{r^2}{r - 2M} {\dot H}_0 - \frac{M}{r-2M}H_1 - r H'_1 \right]\dot{r}_p^2
\]
\begin{equation}
- 3 H'_0 \dot{r}_p - \frac{3}{2}{\dot H}_0 + \frac{3M}{r^2}H_1 
\end{equation}

\begin{table}[totgeo]
\caption{Classification of geodesic terms.}
\vskip 10pt
\small\rm
{\begin{tabular}{c c c c c c}
$\Gamma_{rr}^t \left(\displaystyle\frac{dr}{dt}\right)^3$ &
$2\Gamma_{tr}^t \left(\displaystyle\frac{dr}{dt}\right)^2$ &
$-\Gamma_{rr}^r\left(\displaystyle\frac{dr}{dt}\right)^2$ &
$\Gamma_{tt}^t \left(\displaystyle\frac{dr}{dt}\right)$ &
$-2\Gamma_{tr}^r\left(\displaystyle\frac{dr}{dt}\right)$ &
$- \Gamma_{tt}^r$ \\[12pt] \hline \\
&
$\eta^{tt}\eta_{tt,r}\dot{r}_p^2$ & 
$- \frac{1}{2}\eta^{rr}\eta_{rr,r}\dot{r}_p^2$ &
&
&
$\frac{1}{2}\eta^{rr}\eta_{tt,r}$ \\ [6pt]
\multicolumn{6}{c}{\em Unperturbed Schwarzschild $\alpha_0$ term}\\[3pt] \hline \\
&
$\eta^{tt}_{,r}\eta_{tt,r}\dot{r}_p^2$&
$ - \frac{1}{2}\eta^{rr}_{,r}\eta_{rr,r}\dot{r}_p^2$ &
&
&
$ \frac{1}{2}\eta^{rr}_{,r}\eta_{tt,r}$ \\ [6pt]
&
$\eta^{tt}\eta_{tt,rr}\dot{r}_p^2$ &
$ - \frac{1}{2}\eta^{rr}\eta_{rr,rr}\dot{r}_p^2$ & 
&
&
$ \frac{1}{2}\eta^{rr}\eta_{tt,rr}$\\ [6pt]
\multicolumn{6}{c}{\em $\alpha_1$ term }\\[4pt] \hline \\
&
$2\eta^{tt}\eta_{tt,r}\dot{r}_p$ &
$ - \eta^{rr}\eta_{r,r}\dot{r}_p$ &
&
& \\[4pt]
\multicolumn{6}{c}{\em $\alpha_2$ term }\\[3pt] \hline \\
&
$\eta^{tt}_{,r}\eta_{tt,rr}\dot{r}_p^2$&
$ - \frac{1}{2}\eta^{rr}_{,r}\eta_{rr,rr}\dot{r}_p^2$&
&
&
$ \frac{1}{2}\eta^{rr}_{,r}\eta_{tt,rr}$ \\[3pt]
\multicolumn{6}{c}{\em $\alpha_3$ term }\\[3pt] \hline \\
&
$\eta^{tt}\eta_{tt,r}$&
$ -\frac{1}{2}\eta^{rr}\eta_{rr,r}$&
&
& 
\\[3pt]
\multicolumn{6}{c}{\em $\alpha_4$ term}\\[3pt] \hline \\
&
$2\eta^{tt}_{,r}\eta_{tt,r}\dot{r}_p$&
$-\eta^{rr}_{,r}\eta_{rr,r}\dot{r}_p$&
&
& 
\\[6pt]
&
$2\eta^{tt}\eta_{tt,rr}\dot{r}_p $&
$-\eta^{rr}\eta_{rr,rr}\dot{r}_p $&
&
& 
\\[3pt]
\multicolumn{6}{c}{\em $\alpha_5$ term }\\[3pt] \hline \\
$\eta^{tt}h_{tr,r}\dot{r}_p^3$&  
$- h^{tt}\eta_{tt,r}\dot{r}_p^2$&
$\frac{1}{2}h^{rr}\eta_{rr,r}\dot{r}_p^2$&
$\frac{1}{2}\eta^{tt}h_{tt,t}\dot{r}_p$&   
$ - \eta^{rr}h_{rr,t}\dot{r}_p$& 
$ - \eta^{rr}h_{rt,t}$ \\[6pt]
$-\frac{1}{2}\eta^{tt}h_{rr,t}\dot{r}_p^3$&  
$\eta^{tt}h_{tt,r}\dot{r}_p^2$& 
$- \frac{1}{2}\eta^{rr}h_{rr,r}\dot{r}_p^2$& 
$\frac{1}{2}h^{tr}\eta_{tt,r}\dot{r}_p$&  
$h^{rt}\eta_{tt,r}\dot{r}_p$&  
$- \frac{1}{2}h^{rr}\eta_{tt,r}$ \\[6pt] 
$- \frac{1}{2}h^{tr}\eta_{rr,r}\dot{r}_p^3$ &
&
&
&
&
$\frac{1}{2}\eta^{rr}h_{tt,r}$ \\[3pt] 
\multicolumn{6}{c}{\em $\alpha_6$ term }\\[3pt] \hline \\
$\eta^{tt}_{,r}h_{tr,r}\dot{r}_p^3$&  
$-h^{tt}_{,r}\eta_{tt,r}\dot{r}_p^2$& 
$\frac{1}{2}h^{rr}_{,r}\eta_{rr,r}\dot{r}_p^2$&
$\frac{1}{2}\eta^{tt}_{,r}h_{tt,t}\dot{r}_p$&
$-\eta^{rr}_{,r}h_{rr,t}\dot{r}_p$&
$-\eta^{rr}_{,r}h_{rt,t}$ \\[6pt]
$\eta^{tt}h_{tr,rr}\dot{r}_p^3$&  
$\eta^{tt}_{,r}h_{tt,r}\dot{r}_p^2$ &
$-\frac{1}{2}\eta^{rr}_{,r}h_{rr,r}\dot{r}_p^2$& 
$\frac{1}{2}\eta^{tt}h_{tt,tr}\dot{r}_p$& 
$-\eta^{rr}h_{rr,tr}\dot{r}_p$&
$-\eta^{rr}h_{rt,tr}$ \\[6pt] 
$- \frac{1}{2}\eta^{tt}_{,r}h_{rr,t}\dot{r}_p^3$& 
$-h^{tt}\eta_{tt,rr}\dot{r}_p^2$&
$ \frac{1}{2}h^{rr}\eta_{rr,rr}\dot{r}_p^2$& 
$\frac{1}{2}h^{tr}_{,r}\eta_{tt,r}\dot{r}_p$& 
$ h^{rt}_{,r}\eta_{tt,r}\dot{r}_p$&
$-\frac{1}{2}h^{rr}_{,r}\eta_{tt,r}$ \\[6pt]
$ -\frac{1}{2}\eta^{tt}h_{rr,tr}\dot{r}_p^3$& 
$\eta^{tt}h_{tt,rr}\dot{r}_p^2$&    
$- \frac{1}{2}\eta^{rr}h_{rr,rr}\dot{r}_p^2$&
$\frac{1}{2}h^{tr}\eta_{tt,rr}\dot{r}_p$& 
$h^{rt}\eta_{tt,rr}\dot{r}_p$&
$\frac{1}{2}\eta^{rr}_{,r}h_{tt,r}$ \\[6pt]
$- \frac{1}{2}h^{tr}_{,r}\eta_{rr,r}\dot{r}_p^3$& 
&
&
&
&
$-\frac{1}{2}h^{rr}\eta_{tt,rr}$ \\[6pt] 
$-\frac{1}{2}h^{tr}\eta_{rr,rr}\dot{r}_p^3$& 
&
&
&
&
$\frac{1}{2}\eta^{rr}h_{tt,rr}$ \\[3pt] 
\multicolumn{6}{c}{\em $\alpha_7$ term }\\[3pt] \hline \\
$3\eta^{tt}h_{tr,r}\dot{r}_p^2$ &
$-2h^{tt}\eta_{tt,r}\dot{r}_p$&
$h^{rr}\eta_{rr,r}\dot{r}_p$&
$\frac{1}{2}\eta^{tt}h_{tt,t}$&
$-\eta^{rr}h_{rr,t}$ & \\[6pt]
$- \frac{3}{2}\eta^{tt}h_{rr,t}\dot{r}_p^2$&
$2\eta^{tt}h_{tt,r}\dot{r}_p$&
$-\eta^{rr}h_{rr,r}\dot{r}_p$& 
$\frac{1}{2}h^{tr}\eta_{tt,r}$&
$h^{rt}\eta_{tt,r}$ & \\[6pt]
$-\frac{3}{2}h^{tr}\eta_{rr,r}\dot{r}_p^2 $& 
&
&
& 
& \\[3pt]
\multicolumn{6}{c}{\em $\alpha_8$ term}\\[3pt] 
\end{tabular}}
\end{table}


\section{References}

\begin{thebibliography}{99}


\bibitem{lisa}
http://sci.esa.int/science-e/www/area/index.cfm?fareaid=54;  
http://lisa.jpl.nasa.gov

\bibitem{ghezetal00}
Ghez A Morris M Becklin E E Kremenek T and Tanner A 2000 {\it Nature} {\bf 407} 349

\bibitem{pripul94}
Price R H and Pullin J 1994 {\it Phys.Rev. Lett.} {\bf 72} 3297

\bibitem{vili}
http://www.virgo.infn.it; http://www.ligo.caltech.edu

\bibitem{bakcamlou02}
Baker M Campanelli M and Lousto C 2002 {\it Phys. Rev. D} {\bf 65} 044001

\bibitem{hug01}
Hughes S A 2001 {\it Class Quant Grav} {\bf 18} 4067

\bibitem{mon74}
Moncrief V 1974 {\it Ann. Phys. (N.Y.)} {\bf 88} 323 

\bibitem{rw57}
Regge T and Wheeler J A 1957 {\it Phys. Rev.} {\bf 108} 1063  

\bibitem{zeri}
Zerilli F J 1970 {\it Phys. Rev. Lett.} {\bf 24} 737; 
Zerilli F J 1970 {\it J. Math. Phys.} {\bf 11} 2203;
Zerilli F J 1970 {\it Phys. Rev. D} {\bf 2} 2141; Zerilli F J 1975 {\it Black holes, gravitational waves and cosmology: an introduction to current research} Eds M J Martin R Ruffini J A Wheeler (New York: Gordon \& Breach Science) A-7 
errata corrige


\bibitem{garpri00}
Garat A and Price R H 2000 {\it Phys. Rev. D} {\bf 61} 044006

\bibitem{glniprpu96}
Gleiser R J Nicasio C 0 Price R H and Pullin J 1996 {\it Class. Quantum Grav.} {\bf 13} L117 

\bibitem{sf}
Mino Y Sasaki M and Tanaka T 1997 {\it Phys. Rev. D} {\bf 55} 3457; Quinn T C and Wald R M 1997 {\it Phys. Rev. D} {\bf 56} 3381

\bibitem{as}
Aoudia S Spallicci A 2004 {\it in preparation}

\bibitem{spa99a}
Spallicci A 1999 {\it 
  8$^{th}$ Marcel Grossmann Meeting} 22-28 June 1997 Jerusalem Eds T Piran R Ruffini (Singapore:
World Scientific) 1107  

\bibitem{spa99b}
Spallicci A 1999 {\it 2$^{nd}$ Amaldi Conf. on Gravitational Waves} 1-4 July 1997 Cern Geneve 
Eds E Coccia G Veneziano G Pizzella (Singapore: World Scientific) 303

\bibitem{spa00}
Spallicci A 2000 {\it Recent developments in general relativity} 13$^{th}$ It. Conf. Gen. Rel. \& Grav. Phys.  
21-25 September 1998 Monopoli Eds B Casciaro D Fortunato M Francaviglia A Masiello (Milano: Springer) 371

\bibitem{lo00}
Lousto C O 2000 {\it Phys. Rev. Lett.} {\bf 84} 5251

\bibitem{lo01a}
Lousto C O 2001 {\it Class. Quantum Grav.} {\bf 18} 3989 

\bibitem{vin04}
Vinet J Y Spallicci A 2004 {\it in preparation}

\bibitem{lo01b}
Lousto C O 2001b http://www.aei-potsdam.mpg.de/~lousto/CAPRA/PROCEEDINGS

\bibitem{balo02}
Barack L and Lousto C O 2002 {\it Phys. Rev. D} {\bf 66} 061502

\bibitem{ao03}
Aoudia S 2003 {\it Rapport de stage D.E.A.} Univ. de Nice-Sophia Antipolis 

\bibitem{ri59}
Riemann B 1859 {\it Mon. Not. Berlin Akad.} 671 

\bibitem{hu82}
Hurwitz A 1882 {\it Z. Math. Phys.} {\bf 27} 97
\end{thebibliography}
\end{document}